\documentclass[sigconf,natbib=true%
]{acmart}

\AtBeginDocument{%
  \providecommand\BibTeX{{%
    \normalfont B\kern-0.5em{\scshape i\kern-0.25em b}\kern-0.8em\TeX}}}

\setcopyright{none}
\renewcommand\footnotetextcopyrightpermission[1]{}
\usepackage{booktabs}
\usepackage{graphicx}
\usepackage{enumitem}

\usepackage{lipsum}
\usepackage{float}
\usepackage{subfigure}
\usepackage{multirow,makecell}
\usepackage{pifont}
\usepackage{caption}
\usepackage{balance}
\usepackage{todonotes}
\usepackage{adjustbox}
\usepackage[draft,commandnameprefix=ifneeded]{changes}

\newcommand{\AK}[1]{\textcolor{blue}{{\bf [AK: }{\bf ]}}}
\newcommand{\EK}[1]{\textcolor{green}{{\bf [EK: }{\bf ]}}}

\begin{document}
\fancyhead{}
\title{Corpus-informed Retrieval Augmented Generation of Clarifying Questions}

\author{Antonios Minas Krasakis}
\email{a.m.krasakis@uva.nl}
\affiliation{%
  \institution{University of Amsterdam}
  \country{The Netherlands}
}
\author{Andrew Yates}
\email{a.c.yates@uva.nl}
\affiliation{%
  \institution{University of Amsterdam}
  \country{The Netherlands}
}
\author{Evangelos Kanoulas}
\email{e.kanoulas@uva.nl}
\affiliation{%
  \institution{University of Amsterdam}
  \country{The Netherlands}
}

\renewcommand{\shortauthors}{Krasakis, et al.}

\begin{abstract}

This study aims to develop models that generate corpus informed clarifying questions for web search, in a way that ensures the questions align with the available information in the retrieval corpus. 
We demonstrate the effectiveness of Retrieval Augmented Language Models (RAG) in this process, emphasising their ability to (i) jointly model the user query and retrieval corpus to pinpoint the uncertainty and ask for clarifications end-to-end and (ii) model more evidence documents, which can be used towards increasing the breadth of the questions asked. 
However, we observe that in current datasets search intents are largely unsupported by the corpus, which is problematic both for training and evaluation. 
This causes question generation models to ``hallucinate'', ie. suggest intents that are not in the corpus, which can have detrimental effects in performance.
To address this, we propose dataset augmentation methods that align the ground truth clarifications with the retrieval corpus. 
Additionally, we explore techniques to enhance the relevance of the evidence pool during inference, but find that identifying ground truth intents within the corpus remains challenging.
Our analysis suggests that this challenge is partly due to the bias of current datasets towards clarification taxonomies and calls for data that can support generating corpus-informed clarifications.

\end{abstract}

\maketitle

\section{Introduction}

Open-domain search and Question Answering (QA) systems make the best effort to respond to any user's question or query.
Recent work attempts to quantify the uncertainty in the question-answering models in order to defer from answering a question they are uncertain about~\cite{chen2023quantifying,yin-etal-2023-large}. In a parallel line of research, limited work in open-domain conversational and ad-hoc search systems has investigated enabling them to ask Clarifying Questions (CQs)~\cite{aliannejadi2019asking,min2020ambigqa,zamani2020mimics}. %
The majority of this work employs static models of clarifying question generation, i.e. models that generate a clarifying question independent of the ability of the system to locate the right answer in the underlying corpus, or the potential different answers present therein~\cite{zamani2020generating,hashemi2021learning,hashemi2022stochastic,zhao2022generating,aliannejadi2019asking}.

In this work, we emphasize the importance of generating \textit{corpus-informed} clarifying questions, that are \textit{dynamic} with respect to the collection. We argue that it is critical to defer the ambiguity of the user query by modeling the aspects\footnote{Intents, facets and aspects are used interchangeably in this paper.} of it in the document collection. Since the primary goal of search is to retrieve relevant information, clarifying questions should be generated as a function of the corpus and the relevant information therein.
Failing to be dynamic and corpus-informed poses a risk to user experience, (a) due to the disruption caused by asking generic questions that offer no relevant information, and more severely (b) due to "hallucinations" while generating clarifying questions, ie. presenting users with options or facets that do not exist in the collection.

Some research explicitly models the underlying corpus by relying on pipeline methods that (i) extract keywords or features from the document collection and (ii) generate questions based on these features~\cite{sekulic2021towards,sekulic2022exploiting,zhao2023improving,zhao2022generating,wang2023zero,zhang2021diverse}. However, this separation prevents the joint modeling of the query and its ambiguity, the information in the corpus and the various aspects present therein, and the clarification questions. In this work we advocate for Retrieval Augmented Generation of Clarifying Questions, in a way that jointly models queries and the retrieval corpus to generate questions.
Another line of work achieves state-of-the-art performance using text generation models without intermediate feature extraction~\cite{samarinas2022revisiting,ni2023comparative}, but is limited by the size of the retrieval pool. 
This is a crucial factor, as it determines the breadth of the possible clarifications.

Since the generation model cannot read the entire corpus, Retrieval Augmentation helps to inform the generator of the possible information needs present in the collection. Hence, along with the generator, a retriever is responsible for obtaining a representative sample that covers the uncertainty of the user's query in the collection.
To apply Retrieval Augmentation effectively, it is crucial to (a) select a sample of the corpus that is representative of users' potential information needs, and (b) account for as many relevant documents as possible to cover those. 
For this reason, we propose using \textit{Fusion-in-Decoders (FiD)}, a family of models~\cite{izacard2020leveraging} that are computationally efficient in modelling multiple evidence documents. 
We demonstrate their effectiveness in simultaneously modeling the collection (user queries and retrieved documents) and generating questions, in a way that eliminates the need for an intermediate step of facet or keyword extraction and increases the ability
to model larger parts of the corpus when generating questions.

Furthermore, to ensure dynamic and adaptive nature of clarifying question generation and avoid "hallucinations", we enhance the training setup of question generation models. We find that current datasets~\cite{zamani2020generating} suffer from a disconnect between document evidence gathered from search engine results pages and the ground truth clarification questions derived from user reformulations, negatively impacting the question generator. In light of this, we further explore the relationship between question generation and evidence documents. We find that aligning evidence and generation during training is critical towards preventing "hallucinations"~\cite{krishna2021hurdles} and ensuring the generated question remains faithful to evidence documents, ie. the retrieval corpus.

Having trained an effective and grounded question generator, our focus shifts to retrieving evidence during inference. 
To ensure evidence documents encompass all existing information facets of the original queries, we experiment with inducing novelty in evidence documents to improve facet generation. 
Yet, we find that capturing the ground truth facets present in current datasets with such methods remains a challenge.

Our main research question is \emph{how can we train an end-to-end system that explicitly models the retrieval corpus and generates Corpus-informed clarifying questions?} 
Specifically, we aim to answer the following research questions:
\begin{enumerate}[label=\textbf{RQ\arabic*},leftmargin=*]
    \item\label{rq:rag-cqgen} Can we generate Clarifying Questions end-to-end with Retrieval Augmented Generation models?
    \item\label{rq:optimal-evidence-set} What is the optimal evidence set for training and evaluating that support \textit{Corpus-informed} Retrieval-Augmented Generators of clarifying questions?
    \item\label{rq:retrievak-diversification} How can we enhance the evidence set at test time to assist clarifying question generation?
\end{enumerate}

In particular, we make the following contributions to the research of asking clarifying questions: (a) we define the desired properties of clarifying questions to be tightly dependent on both user queries, as well as documents retrieved for those queries; (b) to this end we develop a dataset that allows for training faithful and adaptive clarifying question models; (c) we propose using Retrieval Augmented Generation for generating clarifying questions end-to-end and show that Fusion-in-Decoder is an effective approach on this task, and (d) we investigate how to enhance search results for the purpose of better informing the generation of clarifying questions. 

\section{Related Work}

We first discuss datasets used in clarifying question generation and evaluation (Section \ref{sec:rw-cq-tasks}), followed by methods for generating clarifying questions with a focus on web search (Section \ref{sec:rw-cq-methods}).

\subsection{Clarifying question datasets}\label{sec:rw-cq-tasks}

Recent work on clarifying questions differs on several aspects, including the type of questions ask, the way presented to the user, and the evaluation methodology for automatic question generation. Below we discuss several types of clarifying questions datasets.%

Clarifying Questions for Web Seach were first introduced in Qulac~\cite{aliannejadi2019asking}, an open-domain information-seeking conversational search dataset. This work builds upon faceted or ambiguous queries from the TREC Web Track dataset~\cite{collins2014trec,clarke2009overview}. 
The authors crowd-sourced clarifying questions to be asked from a search system to the user, as well as answers to these questions from users with different intents. The quality of automatically generated clarifying question was evaluated by the relevance of the retrieved documents.
ClariQ~\cite{aliannejadi2020convai3} extended Qulac with additional topics and built synthetic multi-turn conversations from single-turn clarifications.

MIMICS introduces a clarification pane on the search engine result page (SERP)~\cite{zamani2020generating,zamani2020mimics}. 
This pane includes a template-based question (eg. ``Who are you shopping for'') with multiple candidate answers or search intents (eg. "men, women, kids"). 
Intents are extracted from query reformulation logs and a defined taxonomy. 
The offline evaluation framework compares ground truth intents to the generated ones~\cite{hashemi2021learning,samarinas2022revisiting,wang2021template}, while later work also introduces an online evaluation setup based on user interactions~\cite{tavakoli2022mimics,tavakoli2022analyzing,zamani2020analyzing}.

Another line of research investigates clarification in community forums like StackExchange~\cite{qu2018analyzing,tavakoli2022analyzing,rao2018learning,kumar2020clarq,penha2019introducing}. In these datasets, clarifying questions are written by expert users and the task is often defined as clarifying question retrieval from a pool of questions. 
Last, there are efforts to create clarifying questions on the product search domain, either towards assisting users in product search~\cite{zou2020towards,zhang2018towards}, but also to clarify ambiguities originating from product descriptions~\cite{rao2019answer,zhang2021diverse}. These works differ from our line of work, since they often revolve around structured product metadata. 

For a more comprehensive overview on clarifying questions datasets we refer readers to \citet{rahmani2023survey}.

\subsection{Clarifying questions generation for web search}\label{sec:rw-cq-methods}
A number of approaches have been proposed for generating clarifying questions for web search results. 
Broadly, these works use a combination of rule-based systems, keyword extraction or topic modelling approaches, and Large Language Models (LLMs). 
\citet{hashemi2021learning} proposed NMIR, a transformer architecture that learns multiple intent representations for web queries by matching different document clusters to query intents. 
Later works found that a transformer encoder-decoder approach based on the BART model can outperform NMIR~\cite{samarinas2022revisiting}.
Other research tried combining LLMs with facet extraction methods. \citet{sekulic2021towards} constructed the ClariQ-FKW (Facet KeyWords) dataset and used it to guide GPT-2 in generating clarifications. They find that using facet keywords to guide the LLM helped with grounding and generating useful questions. 
In follow-up work, \citet{sekulic2022exploiting} try to improve upon the facet/keyword extraction part by using part-of-speech tags, entities and LDA topics. They combine those approaches by ranking their output using entropy-based methods and generate a template-based question using the top ranked keyword. 
In contrast to our work, (a) these methods only generate questions addressing one facet and (b) keep the facet extraction part disconnected from question generation.
\citet{samarinas2022revisiting} combines various facet extraction methods, such as autoregressive generation, sequence tagging, extreme multi-labelling classification and LLM prompting in an ensemble, concluding that these methods are often complementary.

Another line of work tries to inform the CQ generation using descriptions of queries and lists of attributes from retrieved web pages~\cite{zhao2022generating}.
Those are extracted with heuristics and filtering, and ranked using learning-to-rank. The top ranked are given to a seq2seq model (QLM~\cite{zamani2020generating}) to generate the final clarifying question. 
Similarly, \citet{zhao2023improving} focuses on complementing web-search results with relational information (eg. ``Windows'' $\rightarrow$ Operating system) extracted from web search results to inform the generation. 
\citet{wang2023zero} introduces another approach that uses LLMs and extracted keywords to generate clarifications. Instead of prompting, they use Neurologic decoding, a constrained generation approach that biases the generation towards these keywords. Mltiple clarifications are generated and then ranked with a CQ ranking system.
In contrast to those our work focuses on modelling the collection and generating a question in an end-to-end way, without the need for intermediate steps.
Finally, another line of work focuses on ranking clarifying questions from a large pool of existing questions~\cite{ou2020clarifying,aliannejadi2020convai3}. However, such question pools do not exist in an open-domain setting and we do not discuss those further.
Those approaches include many different components (eg. keyword extraction, generation) that work in a disconnected way from each other.

Last there is work that focuses on enhancing the training of autoregressive models for generating clarification facets, which are by nature unordered. Next token prediction objectives could unfairly penalize the models for predicting facets in a different ordering. 
To tackle this, \citet{hashemi2022stochastic} generate all possible permutations of facets and backpropagate the minimum loss during training. \citet{ni2023comparative} does a thorough comparison of training objectives and conclude that training on one-facet-at-a-time basis increases performance but hurts facet diversity. This work is orthogonal to ours and can also be applied to our models.

\section{Methodology}
\paragraph{Problem definition.} 
In this work we investigate the problem of clarifying question generation for Web Search, where clarifying questions are presented through a clarification pane in the Search Engine Result Page (SERP). 
In this setting, the user has a search intent $g_i \in G$ and searches for a set of relevant documents $R_i$ in a corpus of documents $C$. 
To express his information need $g_i$, the user issues an ambiguous or faceted query $q$ to the search engine. 
The SERP presents the top ranking results along with a clarification pane that tries to clarify the ambiguity of $q$. 
The clarification pane consists of a clarifying question $CQ$ and some potential answers $F$ to this clarifying question $CQ$. Those answers are basically possible search intents, facets or aspects of query $q$. 

The goal of asking a good clarification question becomes directly related to finding the possible search intents (facets) $F$ that query $q$ engulfs. 
To find such search intents $F$, it is important to take into account the document collection $C$ the user searches through. In failing to do so, we risk presenting search engine users with facets that are not found in $C$. 

\paragraph{RAG models.} Retrieval Augmented Generation models have become the standard choice for Knowledge-Intensive tasks where generation is required. Their strength lies on combining LLMs, that are proficient in text generation, with retrieval, that allows them to access information from a large non-parametric memory like a document collection~\cite{asai2023retrieval}.
That also motivates their use for our task, end-to-end and Corpus-informed clarifying question generation. 
Specifically, we use a Fusion-in-Decoder (\textit{FiD}) model~\cite{izacard2020leveraging} as a question generator, due to its efficiency and effectiveness. 
\textit{FiD} is an encoder-decoder model, but its encoder models input documents independently (no cross-attentions), producing individual embeddings. The decoder fuses the information from those embedding, generating an output answer. Due to the lack of cross-attentions between documents, \textit{FiD} models can model longer context, ie. multiple retrieved documents with the same GPU memory requirements. 

\paragraph{Retrieval.} We experiment with various types of retrieval techniques, such as lexical ($BM25$), semantic ($Contriever$), as well as the documents that originated from the BING SERP provided with the MIMICS dataset ($BING$).
For $BM25$ search, we use the Pyserini toolkit~\cite{lin2021pyserini} using the default parameters. 
For semantic retrieval, we use the Contriever architecture that has been jointly pretrained with the $Atlas$ checkpoints, but observe in our preliminary experiments that retrieval performance of the retriever jointly pretrained with $Atlas$ models is sub-optimal. 
To this end, we initialize the retriever from the unsupervised pretraining of Contriever ($Contriever$) and the checkpoint finetuned on MSMarco\cite{nguyen2016ms} ($Contriever-FT$)\footnote{https://huggingface.co/facebook/contriever\\https://huggingface.co/facebook/contriever-msmarco}
Additionally, we consider two retrieval variants, namely non-aligned ($*|Q$) or facet-aligned ($*|Q,F$). For the former, we use only the original user query to retrieve, while in case of the latter we do multiple retrieval rounds using the query and each facet and interleave the retrieved documents (without replacing duplicates).

When experimenting with novelty methods (Section \ref{sec:results-rq3}), we train the retriever with knowledge distillation from the question generator~\cite{izacard2020distilling}. 
In practice, this method uses attention scores of document embeddings in the decoder as a proxy for document relevance. In Question Answering, this signal corresponds to detecting relevant passages for a question, while in our setting it promotes novel documents that contain the ground truth facets.

\begin{table*}[t]

\begin{adjustbox}{width=\textwidth, center}
    \begin{tabular}{lr|rrr|rrr|rrr|rrrr}
    \toprule
     Model & Co-Gen & \multicolumn{3}{c}{Term Overlap} & \multicolumn{3}{c}{Exact Match} & \multicolumn{3}{c}{Set-BERT} & \multicolumn{4}{c}{Set-BLEU} \\
     & & Prec. & Rec. & F1 & Prec. & Rec. & F1 & Prec. & Rec. & F1& BLEU1 & BLEU2 & BLEU3 & BLEU4 \\
    \midrule

    \multicolumn{15}{c}{Our Models (FiD)}\\\midrule

    FiD-CQGen & \ding{52} & \textbf{0.3459} & 0.3041  & \textbf{0.3095}  & 0.0781  & 0.0593  & 0.0645  & 0.4232  & 0.4507  & 0.4323  & 0.3815  & 0.3265  & 0.2969  & 0.2777  \\

    FiD-RevGen ${^a}$ & \ding{52} & \underline{0.3418} $^{b,c}$ & 0.3061  & \underline{0.3093} $^{c}$ & \textbf{0.0854}  $^{c}$ & 0.0677  & \textbf{0.0723}  $^{c}$ & \underline{0.4322}  $^{c}$ & \underline{0.4571}  $^{b}$ & \underline{0.4400}  $^{b}$ & \underline{0.3858}  $^{c}$ & \underline{0.3310} $^{c}$  & \textbf{0.3020} $^{c}$ & \textbf{0.2826}  $^{c}$ \\

    FiD-AspGen & \ding{55} & 0.3416  & 0.2998  & 0.3061  & \underline{0.0829}  & 0.0613  & 0.0671  & 0.4217  & 0.4466  & 0.4296  & 0.3762  & 0.3224  & 0.2937  & 0.2754  \\
    \midrule
    
    \multicolumn{15}{c}{Baselines}\\\midrule

    $BART(query+docs)$ ~\cite{samarinas2022revisiting} $^{b}$ & \ding{55} & 0.3003 & \underline{0.3199} $^{a}$ & 0.2984 & 0.0770 & \underline{0.0694} & \underline{0.0711} & \textbf{0.4452} $^{a}$ & \textbf{0.4673}  & \textbf{0.4516}  & \textbf{0.3994} $^{a}$ & \textbf{0.3355} & \underline{0.3000}  & \underline{0.2778}\\

    $Faspects$ ensemble~\cite{samarinas2022revisiting} $^{c}$  & \ding{55} & 0.2252 & \textbf{0.3296} $^{a}$ & 0.2588 & 0.0499 & \textbf{0.0795} & 0.0596 & 0.3558 & 0.3475 & 0.3473 & 0.2625 &  0.2076 & 0.1826 & 0.1681 \\
    
    \midrule

    \multicolumn{15}{c}{Oracles/Ablations}\\\midrule

    $BART-closedbook$  & \ding{55} & 0.0779 & 0.0530 & 0.0591 & 0.0294 & 0.0226 & 0.0245 & 0.3706 & 0.3071 & 0.3312 & 0.1280 & 0.0632 & 0.0396 & 0.0333 \\

    FiD-AspGen-$closedbook*$ & \ding{55} & 0.2973  & 0.2499  & 0.2616  & 0.0293  & 0.0202  & 0.0227  & 0.3956  & 0.4188  & 0.4034  & 0.3448  & 0.2927  & 0.2648  & 0.2480  \\
    FiD-AspGen-$oracle-compressed$ & \ding{55} & 0.9968  & 0.9138  & 0.9480  & 0.9867  & 0.8916  & 0.9296  & 0.9001  & 0.8997  & 0.8999  & 0.8991  & 0.8988  & 0.8965  & 0.8825 \\
    FiD-AspGen-$oracle$ & \ding{55} & 0.9999  & 0.9999  & 0.9999  & 0.9999  & 0.9999  & 0.9999  & 1.0000  & 1.0000  & 1.0000  & 1.0000  & 1.0000  & 0.9974  & 0.9810  \\
    \midrule

    \end{tabular}
\end{adjustbox}
\caption{Aspect generation performance when using bing-snippets~\cite{zamani2020mimics} as evidence documents. Co-Gen stands for co-generating the template-based questions and target facets. Bold designates the top system while underline the second top excluding the oracle systems. 
Superscripts indicate statistically significant improvements wrt. FiD-RevGen $^{a}$, BART $^{b}$ and the Faspects ensemble $^{c}$ (paired T-test, $p-value<0.05$ and Bonferroni correction for multiple hypothesis testing).
}
\label{tab:main-table}
\end{table*}

\section{Experimental Setup}
In this section we outline our experimental setup.

\subsection{Datasets and Evaluation}
As discussed in Section \ref{sec:rw-cq-tasks}, a number of Clarifying Question datasets and setups exist.
In this work, we focus on Web Search Clarifying Questions and perform our experiments on the MIMICS dataset. 
Our choice is based on a number of factors, namely: (a) the presentation of clarifying questions that happens via a clarification pane on a SERP. In our view, this presentation is less intrusive than a conversational system that interrupts the user's journey without presenting results to ask a question, and hence less likely to harm user experience \cite{avula2022effects}, and (b) a more straightforward and robust evaluation method that directly evaluates generated facets~\cite{krasakis2020analysing}. 
In contrast, the QuLaC~\cite{aliannejadi2019asking} evaluation framework involves a user answering the clarifying question and ranking documents to judge question quality. 
However, evaluation can be particularly noisy, since both of these parts are challenging. User answers largely depend on their cooperativeness~\cite{sekulic2022evaluating}, while ranking documents with clarification-based queries is particularly challenging~\cite{krasakis2020analysing}.

Following prior work on $MIMICS$, we train on $MIMICS-Click$ and test on $MIMICS-Manual$~\cite{samarinas2022revisiting,hashemi2021learning,hashemi2022stochastic}.

\subsubsection*{Evaluation metrics}
Following previous research~\cite{hashemi2022stochastic,hashemi2021learning,samarinas2022revisiting,zhao2023improving}, we focus on a number of lexical and semantic metrics that measure overlap of generated facets to the ground truth ones. 
Given a sequence of ground truth facets, $G = g_1, g_2, g_n$, and a sequence of generated facets, $F = f_1, f_2, ..., f_m$, we assess aspect generation quality using: 
\begin{itemize}
    \item Term Overlap ($P,R,F1$): measures lexical overlap at the word level, i.e. overlap between words in $G$ and $F$.
    \item Exact Match ($P,R,F1$): measures exact lexical match at the facet level, i.e. whether $f_i = g_j$.
    \item Set-BERT~\cite{hashemi2021learning}: measures semantic overlap at the facet level based on BERT-score~\cite{zhang2019bertscore}%
    \item Set-BLEU-ngram\cite{papineni2002bleu,hashemi2021learning}: measures n-gram overlap at the facet level%
\end{itemize}

For metrics computed on the facet level (Set-BERT and Set-BLEU), we follow prior work and create the best matching pairs between $F$ and $G$ using the BLUE-1 score~\cite{hashemi2021learning}.

\subsection{Baselines}

Due to variations in (a) metric implementation, (b) BERT-Scores' variability to package versions and (c) the use of different test sets \footnote{See footnote 6 of \citet{samarinas2022revisiting}}, 
reported results for a method often differ across papers, indicating a serious reproducibility issue~\cite{samarinas2022revisiting,hashemi2021learning,zhao2023improving}.
To circumvent this issue and ensure a fair comparison, we only compare with publicly available baselines and use the evaluation code of Faspect\cite{samarinas2022revisiting}\footnote{https://github.com/algoprog/Faspect/}. 

We compare with $BART$, a SOTA encoder-decoder aspect extraction model that uses queries and evidence documents to generate facets from \cite{samarinas2022revisiting}. 
In comparison with our \textit{FiD} models, BART is more heavy computationally, since its encoder computes cross-attentions between the evidence documents.
Additionally, we compare with Faspect~\cite{samarinas2022revisiting}, a strong Recall-optimised ensemble model that relies on multiple diverse models for generating facets. Those models include the BART baseline we used here, as well as other Sequence Labelling, Extreme Facet Classification or Unsupervised Facet extraction methods. Ensembling is done using the Round-Robin algorithm~\cite{samarinas2022revisiting}, that practically interleaves results coming from different models until generating the maximum amount of facets ($5$).
We do not compare with baselines that focus on optimising the training objectives for generating set-based predictions, since those are orthogonal to our work and beyond the scope of this paper \cite{ni2023comparative}.

\subsection{Implementation details}
We initialize the question generator (\textit{FiD}) model from an $Atlas-base$ checkpoint, which is pretrained with an unsupervised language modelling objective and exhibits good few-shot abilities~\cite{izacard2022few}\footnote{https://github.com/facebookresearch/atlas/}.
Unless stated differently, we use a training batch size of $32$ to maximize our GPU memory usage, and a maximum generation length of $64$ to fit the longest generation output. The maximum length of the encoded documents is also set to $64$, since our collection consists of small bing snippets and MSMarco-passage.
We do early stopping while optimizing for Exact Match F1 (chosen due to reliability and computational efficiency) on 
a held out validation set, taken from MIMICS-Click.
We perform our experiments on one Nvidia RTX A6000 GPU. 

\section{Results}

In this Section, we discuss our experimental results and the answers to our Research Questions.

\subsection{Can we generate end-to-end Clarifications with Retrieval Augmented Generation?}
\label{sec:results-main-rq}

In this section, we try to answer \ref{rq:rag-cqgen}, that is whether we can generate Clarifying Questions end-to-end with Retrieval Augmented Generation models.
In this part, we use as Retrieval Augmentation (evidence) the BING snippets provided along with the MIMICS dataset~\cite{zamani2020generating}. We compare \textit{Fusion-in-Decoder (FiD)} models with other RAG and closed book models and present results on Table~\ref{tab:main-table}.
Following prior work, we only report facet generation performance so results are comparable across different methods. %

We experiment with three variants of \textit{FiD}, depending on whether the model only predicts facets (\textit{FiD-AspGen}) or jointly generates facets and the template-based questions of MIMICS (\textit{FiD-CQGen}, \textit{FiD-RevGen}). \textit{FiD-CQGen} predicts question and facets in the original order of MIMICS (question followed by the aspects), while \textit{FiD-RevGen} first outputs the aspects.
We find that all variants are competitive, with \textit{FiD-RevGen} performing best across Exact Match metrics while being able to co-generate questions and aspects. For this reason, in the rest of our experiments we use this variant, unless stated differently.

Next we compare \textit{FiD} models with \textit{BART}, the SOTA seq2seq baseline. 
We observe that results are mixed, with significant differences occurring for both models on different metrics. $FiD$ significantly outperforms in Term Overlap Precision metrics, but loses in Term Overlap Recall and Set-BERT Precision, 
Overall, BART seems to be better in terms of recall, and also when compared using the Set-BERT metric, which computes facet-to-facet semantic similarities. 
The Set-BERT metric, although used in relevant literature~\cite{zhang2019bertscore,hashemi2021learning,hashemi2022stochastic,samarinas2022revisiting}, has significant shortcomings, as it favors method that generate a large number of facets.
This is attributed to the fact that our methods generate less facets on average compared to the baselines.
The FiD method is also on par with BART when using Set-BLUE, outperforming BART for larger n-grams. In general, BART achieves higher Recall while FiD is stronger in terms of Precision.
This comparison between \textit{FiD} and \textit{BART} is important, because \textit{FiD} fuses document embeddings in the decoder and therefore can model lengthier or more evidence documents than \textit{BART}, with the same GPU memory footprint. 
The benefits of this efficiency are not visible here and are explored in Section~\ref{sec:increase-context}, due to the evidence snippets being few and very short in length.

Comparing $FiD$ with Faspects, a Recall-optimized ensemble that contains BART, we observe that this trend is magnified further. Faspects produces more facets which at times leads to better Recall, at the cost of much lower metrics in terms of Precision as well as Set-BERT and Set-BLEU measures.

Furthermore, we assess whether FiD can successfully compose answers that depend on multiple evidence documents. 
It is important to assess how the information bottleneck of FiD affects aspect generation performance, because FiD models are widely used on open-domain QA, a task that in contrast to ours does not necessarily require compositionality across evidence documents. 
To do so, we run an oracle experiment where we provide the target facets directly as evidence documents, either in the form of a single evidence document ($oracle$) or in multiple and independently encoded evidence documents ($oracle-compressed$). We observe only a slight drop in performance, and hence we conclude that FiD models can successfully utilise all evidence documents towards the generation.
Lastly, we perform closed-book generation (without using evidence passages), we observe a big drop across performance metrics. By manual inspection, we observe that closed-book models are able to generate facet words for some queries, while they often resort in capturing reoccurring patterns from the training set (eg. on shopping related queries: ``Are you shopping for: $men$, $women$, $kids$ ?'').

Our results in this Section suggest that \textit{RAG} models and in particular \textit{FiD} can be used to effectively and efficiently model the retrieval corpus and generate clarifying questions end-to-end.

Overall, we answer \ref{rq:rag-cqgen} positively and conclude that Retrieval-Augmented-Generation models such as \textit{FiD} are effective in jointly modelling the query, collection and generating clarifying questions in an end-to-end way. 
We find that those \textit{FiD} is competitive with more computationally expensive baselines such as \textit{BART}, suggesting that it can be effectively used to model larger parts of the collection and therefore generate more informed clarifying questions.
Further, we verify the finding of previous works~\cite{samarinas2022revisiting,sekulic2022exploiting} that evidence documents are crucial for generating good clarifying questions and strengthen our motivation towards performing \textit{Corpus-informed Clarifying Question Generation}.

\subsection{Building evidence sets to support Corpus-informed Retrieval-Augmented Generation of Clarifying Questions}
\label{sec:results-alignment}
\begin{table}[]

\begin{adjustbox}{width=\linewidth, center}
\begin{tabular}{ll|cc|r}
\multirow{2}{*}{Collection} & \multirow{2}{*}{Retriever} & Term Overlap  & Exact Match & \multirow{2}{*}{Evidence set type}\\
 & & Recall &  Recall \\\toprule

Bing snippets~\cite{zamani2020generating} & $Bing$ & 0.518 & 0.185 & diversified\\\hline
MSMarco-passage & $BM25|Q$ & 0.448 & 0.151 & non-diversified\\
MSMarco-passage & $Contriever-FT|Q$ & 0.434 & 0.124 & semantic \\\hline 
MSMarco-passage & $BM25|Q,F$ & 0.813 & 0.380 & aligned-lexical \\
MSMarco-passage & $Contriever-FT|Q,F$ & 0.761 & 0.345 & aligned-semantic
\\\hline

\end{tabular}
\end{adjustbox}
\caption{Alignment statistics between evidence documents and target facets in clarifying question}
\label{tab:alignment-stats}
\end{table}
\begin{table*}[t]

\begin{adjustbox}{width=\textwidth, center}
    \begin{tabular}{ll|rrr|rrr|rrr|rrrr}
    \toprule
     Model & evidence (train\&test) & \multicolumn{3}{c}{Term Overlap} & \multicolumn{3}{c}{Exact Match} & \multicolumn{3}{c}{Set-BERT} & \multicolumn{4}{c}{Set-BLEU} \\
     & & Prec. & Rec. & F1 & Prec. & Rec. & F1 & Prec. & Rec. & F1& BLEU1 & BLEU2 & BLEU3 & BLEU4 \\\midrule

    BART  &  [Bing] & 0.3003 & \textbf{0.3199} & 0.2984 & 0.0770 & \textbf{0.0694} & 0.0711 & \textbf{0.4452} & \textbf{0.4673} & \textbf{0.4516} & \textbf{0.3994} & \textbf{0.3355} & 0.3000  & 0.2778 \\ 
    FiD &  [Bing] $^a$ & \textbf{0.3418} $^c$ & 0.3061 $^c$  & \textbf{0.3093} $^c$  &\textbf{ 0.0854}  & 0.0677 & \textbf{0.0723}  & 0.4322  & 0.4571  & 0.4400  & 0.3858  & 0.3310  $^c$ & \textbf{0.3020} $^c$ & \textbf{0.2826}  $^c$ \\

    \hline
    
    FiD & [BM25|Q,F] & 0.4782  & 0.4104  & 0.4220  & 0.1605  & 0.1154  & 0.1290  & 0.4599  & 0.4920  & 0.4700  & 0.4114  & 0.3726  & 0.3504  & 0.3329  \\
    
    FiD & [Contriever|Q,F] & 0.4714  & 0.4338  & 0.4326  & 0.1674  & 0.1296  & 0.1413  & 0.4847  & 0.5197  & 0.4964  & 0.4354  & 0.3926  & 0.3694  & 0.3507  \\
    
    FiD  & [Contriever-FT|Q,F] $^b$ & \textbf{0.5022}$^{a,c}$  & \textbf{0.4517}$^{a,c}$  & \textbf{0.4556}$^{a,c}$  & \textbf{0.1886}$^{a,c}$  & \textbf{0.1497}$^{a,c}$  & \textbf{0.1609}$^{a,c}$  & \textbf{0.4879 }$^{a,c}$ & \textbf{0.5204}$^{a,c}$ & \textbf{0.4980 }$^{a,c}$ & \textbf{0.4376 }$^{a,c}$ & \textbf{0.3990}$^{a,c}$ &\textbf{ 0.3775}$^{a,c}$ &\textbf{ 0.3597}$^{a,c}$  \\
    \hline
    
    FiD & [BM25|Q] & \textbf{0.3216}  & \textbf{0.2912}  & \textbf{0.2928}  & 0.0659  & 0.0540  & 0.0568  & 0.4244  & \textbf{0.4485}  & \textbf{0.4320}  & 0.3754  & \textbf{0.3184}  & \textbf{0.2883}  & \textbf{0.2698}  \\

    FiD & [Contriever|Q] & 0.3095  & 0.2772  & 0.2799  & 0.0544  & 0.0433  & 0.0464  & 0.4218  & 0.4434  & 0.4281  & 0.3688  & 0.3108  & 0.2805  & 0.2625  \\ %
    
    FiD & [Contriever-FT|Q] $^c$ & 0.3202  & 0.2905  & 0.2917  & \textbf{0.0742}  & \textbf{0.0592}  & \textbf{0.0630}  & \textbf{0.4245 } & 0.4462  & 0.4309  & \textbf{0.3754}  & 0.3175  & 0.2868  & 0.2679  \\

    \end{tabular}
\end{adjustbox}
\caption{Effect of different evidence pools in facet generation performance 
(evidence used during training and inference)%
. 
Superscripts $^{a,b,c}$ indicate significant improvements wrt. best model from each category 
(paired T-test, $p-value<0.05$ and Bonferroni correction for multiple hypothesis testing.
Best results across group are \textbf{boldfaced}.
}
\label{tab:evidence-effect-RAG}
\end{table*}

In this section, we focus on the role of retrieval and evidence documents, as those are fundamental to generating good clarifications. 
Our hypothesis here is that if evidence documents and ground truth facets are not aligned during training, that is if ground truth facets do not appear in the evidence documents, then models learn to produce these facets out of their internal model knowledge and eventually stop relying on the retrieved evidence~\cite{krishna2021hurdles}. 
This means that the model becomes corpus-agnostic and static, with a high risk to "hallucinate", ie. present facets to the user that are either irrelevant to her query or do not exist in the retrieval corpus.
Either of these cases would have \textit{detrimental effects for user experience}. This would happen if users are presented with irrelevant facets, or facets that are irretrievable on this search collection $C$, causing the ultimate goal of search (retrieving a relevant document) to fail.
Hence, we emphasize the importance of generating Corpus-informed Clarifying Questions.
Overall, we aim to answer \ref{rq:optimal-evidence-set}, that is how to build evidence sets that can support \textit{Corpus-informed} Clarifying Question generation models. 
Specifically we look into the following questions: (a) does the corpus contain information on all ground truth facets, and can retrieval algorithms bring them up (Section~\ref{sec:alignment}), (b) what is the effect of missing or irretrievable facets in the clarifying question generation (Section~\ref{sec:retrieval-effect-RAG}), (c) do clarifying generation model remain faithful to given evidence documents and how does training affect that (Section~\ref{sec:responsiveness}), and (d) does increasing the document pool help in recovering missing facets~\ref{sec:increase-context})?

\subsubsection{Measuring alignment between target facets and evidence documents} 
\label{sec:alignment}
First, we investigate to what extent the evidence set ($BING$ snippets) we used in Section~\ref{sec:results-main-rq}, and is commonly used in the literature~\cite{hashemi2021learning,hashemi2022stochastic,samarinas2022revisiting}, is aligned with the target ground truth facets we try to generate. 
As a proxy for relevance, we use \textit{Term-Overlap-Recall} that measures how many of all facet words appear in the evidence pool, and \textit{Exact-Match-Recall} that measures whether the entire facet appears verbatim.

In Table~\ref{tab:alignment-stats}, we measure the alignment of the provided \textit{Bing-snippets} as well as evidence pools retrieved using lexical (BM25) and semantic (Contriever) methods. For Bing snippets, we observe that only $50\%$ of facet words appear in the evidence pool. %
When retrieving documents from the MSMarco passage collection~\cite{nguyen2016ms} using the query, alignment is even lower with 43\% and 45\% of the facet words appearing in the evidence set.
Last when we use both the original query and corresponding ground truth facets to construct the evidence pool (by expanding the original query with a single facet and interleaving the top-K results of all facets into a single ranking), the alignment statistics almost double. However, even in this case and when performing lexical retrieval ($BM25|Q,F$), a large number of facet words ($\sim20\%$) are still absent from the evidence pool. This shows that many facets in this dataset might be irretrievable or not even existing, creating a shortcoming in terms of model training (possible hallucinations) and also in terms of evaluation and result reliability. 
\subsubsection{Effect of retriever in generation performance}
\label{sec:retrieval-effect-RAG}

In Table~\ref{tab:evidence-effect-RAG} we use these different evidence pools to explore how they affect downstream performance. 
It is evident that aligning evidence and generation ($*|Q,F$) during training and inference brings a big boost and significant performance improvements across metrics. Most notably the strict Exact Match metric that increases from $\sim6$-$7\%$ to up to $\sim16\%$ when our best retriever ($Contriever-FT$) is used. 
Overall, facet extraction scores follow the trend of the alignment statistics of Table \ref{tab:alignment-stats}. Using only the query ($*|Q$) to retrieve evidence has slightly smaller facet alignment than the Bing snippets and this same trend is reflected in facet extraction performance.

We observe that using better retrievers (ie. the finetuned Contriever) only results in gains in the ($*|Q,F$) setting. This shows that retrievers have to be biased towards the ground truth facets to find them, and suggests the presence of other prevalent facets within the collection. We explore this issue further in Sections \ref{sec:evidence-diversification} and \ref{sec:discussion}.
It is also noteworthy that the generator benefits significantly more from high-quality semantic retrieval ($Contriever-FT$) than lexical ($BM25$), even when measuring lexical metrics. 
This is important because the semantic pool of documents has less lexical overlap with the facets ($0.345$ vs $0.380$, Table \ref{tab:alignment-stats}), yet $FiD$ benefits more from this retriever pool, rather than the lexical one. 

Therefore, we conclude 
that $FiD$ can successfully extract and paraphrase facets from evidence pools without the need to find them verbatim in the evidence, and verify that retrieval quality is of great importance when generating clarifications.

\subsubsection{Faithfulness of Question Generators towards evidence documents}
\label{sec:responsiveness}

\begin{table*}[]

\begin{adjustbox}{width=.8\textwidth, center}
\begin{tabular}{ll|ccc|ccc}
\multirow{2}{*}{Model} &\multirow{2}{*}{Train evidence set} & \multicolumn{3}{c}{Term Overlap}  & \multicolumn{3}{c}{Exact Match} \\
 && $Recall$ & $Recall-LOO$ & $\Delta Recall (\%)$ & $Recall$ & $Recall-LOO$ & $\Delta Recall (\%)$ \\\toprule

Non-Diversified & $BM25|Q$ & 38.138 & 35.987 & -5.64\% & 6.047 & 3.369 & -44.30\% \\

Diversified & Bing snippets~\cite{zamani2020generating} & 
40.233 & 36.700 & -8.78\% & 7.833 & 3.531 &  -54.92\%
\\
Aligned & $BM25| Q,Fß$ &45.804 & 35.247 & -23.05\%  & 14.692 & 3.044 &  -79.28\% \\\bottomrule

\end{tabular}
\end{adjustbox}
\caption{Faithfulness of question models to input evidence documents, using a leave-one-out (LOO) evaluation.}
\label{tab:responsiveness-LOO}
\end{table*}
Next, we test the faithfulness of question generators to the evidence pool. We 
emphasize the importance of this property to ensure clarifications are grounded in the retrieval corpus, where search will take place. 
To probe model faithfulness, we design a leave-one-out (\textit{LOO}) evaluation, where  documents corresponding to a facet are included or dropped from the evidence. 
We start from an evidence pool containing all facets (similar to $*|Q,F$) and measure $Recall$ of a random ground truth facet before or after removing its documents. We probe the generators trained on the Bing snippets evidence set, the non-diversified $BM25|Q$ and the aligned generator $BM25|Q,F$.

As we can see on Table \ref{tab:responsiveness-LOO},  aligned Fusion-in-Decoders are much more likely to capture the facet if the evidence documents contain it, in contrast to the non aligned ones ($Term Overlap$ and $Exact Match$ - $Recall$). $Recall-LOO$ performance is roughly the same across models, with facet recall of aligned models being slightly lower.\footnote{Note that since facets often include query words on mimics (eg. ``\textit{Leiden }zip code'') Term Overlap metrics remain relatively high.}

At the same time, the drop in Recall of the facet that is removed from the evidence set ($\Delta Recall$) is much larger in aligned models, demonstrating substantially more faithful clarifications towards the evidence documents. 
This hints that previous question generators lack sufficient evidence grounding, which is essential for creating \textit{Corpus-informed Clarifications}.

\begin{table*}[t]

\begin{adjustbox}{width=\textwidth, center}
    \begin{tabular}{ll|rrr|rrr|rrr|rrrr}
    \toprule
     Model & train. evidence& \multicolumn{3}{c}{Term Overlap} & \multicolumn{3}{c}{Exact Match} & \multicolumn{3}{c}{Set-BERT} & \multicolumn{4}{c}{Set-BLEU} \\
     & & Prec. & Rec. & F1 & Prec. & Rec. & F1 & Prec. & Rec. & F1& BLEU1 & BLEU2 & BLEU3 & BLEU4 \\\midrule\midrule
    \multicolumn{15}{c}{Test-set evidence set: bing snippets}\\\midrule
     FiD &  [BING] &\textbf{ 0.3418 } & \textbf{0.3061}  & \textbf{0.3093}  & \textbf{0.0854}  & \textbf{0.0677}  & \textbf{0.0723}  & \textbf{0.4325}  & \textbf{0.4574}  & \textbf{0.4403}  & \textbf{0.3857}  & \textbf{0.3308}  & \textbf{0.3018}  & \textbf{0.2826}  \\

     FiD &  [BM25|Q] &  0.3257  & 0.3005  & 0.2993  & 0.0774  & 0.0615  & 0.0652  & 0.4218  & 0.4465  & 0.4296  & 0.3746  & 0.3197  & 0.2904  & 0.2719  \\
     
     FiD & [BM25|Q,F]& 0.3353  & 0.2951  & 0.3015  & 0.0753  & 0.0542  & 0.0595  & 0.4185  & 0.4451  & 0.4269  & 0.3721  & 0.3185  & 0.2897  & 0.2716  \\

      FiD & [Contr-FT|Q,F] & 0.3217  & 0.3012  & 0.2983  & 0.0743  & 0.0604  & 0.0633  & 0.4232  & 0.4494  & 0.4313  & 0.3783  & 0.3212  & 0.2911  & 0.2725  \\
      FiD & [Contr-FT|Q]  & 0.3269  & 0.2998  & 0.2993  & 0.0817  & 0.0665  & 0.0701  & 0.4252  & 0.4483  & 0.4321  & 0.3753  & 0.3201  & 0.2903  & 0.2717  \\
      
      \hline
      
    \multicolumn{15}{c}{Test-set evidence set: non-diversified [$BM25|Q$]}\\\midrule
    FiD &  [Bing] &  \textbf{0.3228  }& 0.2798  & 0.2868  & 0.0571  & 0.0451  & 0.0482  & 0.4022  & 0.4299  & 0.4117  & 0.3634  & 0.3068  & 0.2770  & 0.2584  \\ %
    FiD & [BM25|Q] & 0.3216  & \textbf{0.2912}  & \textbf{0.2928}  & \textbf{0.0659}  & \textbf{0.0540}  & \textbf{0.0568}  & \textbf{0.4244}  & \textbf{0.4485}  & \textbf{0.4320}  & \textbf{0.3751}  & \textbf{0.3183}  & \textbf{0.2883}  & \textbf{0.2698}  \\
    FiD & [BM25|Q,F] & 0.3126  & 0.2661  & 0.2763  & 0.0517  & 0.0386  & 0.0424  & 0.4067  & 0.4314  & 0.4146  & 0.3640  & 0.3060  & 0.2752  & 0.2567  \\\hline %

    \multicolumn{15}{c}{Test-set evidence set: non-diversified [$Contr-FT|Q$]}\\\midrule
    FiD & [Contr-FT|Q] & \textbf{0.3202} & \textbf{0.2905}  & \textbf{0.2917}  & \textbf{0.0742}  & \textbf{0.0592}  & \textbf{0.0630}  & \textbf{0.4245}  & \textbf{0.4462}  & \textbf{0.4309}  & \textbf{0.3754}  & \textbf{0.3175}  & \textbf{0.2868}  & \textbf{0.2679}  \\

    FiD & [Contr-FT|Q,F] & 0.3133  & 0.2778  & 0.2831  & 0.0594  & 0.0477  & 0.0507  & 0.4172  & 0.4402  & 0.4243  & 0.3744  & 0.3149  & 0.2834  & 0.2644  \\

    \end{tabular}
\end{adjustbox}
\caption{Effect of evidence grounding during training (for different evidence sets at inference time) %
}
\label{tab:table-OOD}
\end{table*}
 
\begin{table*}[]

\begin{adjustbox}{width=\textwidth, center}
    \begin{tabular}{ll|rrr|rrr|rrr|rrrr}
    \toprule

     Model & Inference & \multicolumn{3}{c}{Term Overlap} & \multicolumn{3}{c}{Exact Match} & \multicolumn{3}{c}{Set-BERT} & \multicolumn{4}{c}{Set-BLEU} \\
          & evidence pool  & Prec. & Rec. & F1 & Prec. & Rec. & F1 & Prec. & Rec. & F1& BLEU1 & BLEU2 & BLEU3 & BLEU4 \\\midrule

    [Contr-FT|Q] & [Contr-FT|Q] & \textbf{0.3202} & \underline{0.2905}  & \textbf{0.2917}  & \textbf{0.0742}  & \underline{0.0592}  & \underline{0.0630}  & \underline{0.4245}  & \underline{0.4462}  & \underline{0.4309}  & 0.3754  & 0.3175  & \underline{0.2868}  & \underline{0.2679}  \\\hline

    [Contr-FT|Q,F] & MMR ($\lambda=0.5$) &  0.3052  & 0.2763  & 0.2790  & 0.0559  & 0.0458  & 0.0485  & 0.4216  & 0.4441  & 0.4284  & 0.3772  & \underline{0.3179}  & 0.2861  & 0.2670 \\ %
    
    [Contr-FT|Q,F] & MMR ($\lambda=0.7$) & 0.3085  & 0.2767  & 0.2807  & 0.0575  & 0.0471  & 0.0496  & 0.4217  & 0.4444  & 0.4286  & 0.3776  & 0.3174  & 0.2852  & 0.2660 \\ %
    
    [Contr-FT|Q,F] & MMR ($\lambda=0.9$)  & 0.3132  & 0.2784  & 0.2835  & 0.0598  & 0.0484  & 0.0512  & 0.4187  & 0.4416  & 0.4257  & \underline{0.3758}  & 0.3164  & 0.2844  & 0.2653 %
    \\\midrule

     \multicolumn{2}{c}{Contr-FT-Train} & 0.3001  & 0.2715  & 0.2738  & 0.0560  & 0.0451  & 0.0482  & 0.4198  & 0.4416  & 0.4262  & 0.3655  & 0.3070  & 0.2762  & 0.2578 \\ %
     \multicolumn{2}{l}{Contr-FT-Train-WarmedUpQ}  & \underline{0.3150}  & \textbf{0.2935}  & \underline{0.2909}  & \underline{0.0723}  & \textbf{0.0639}  & \textbf{0.0652}  & \textbf{0.4340}  & \textbf{0.4553}  & \textbf{0.4402}  & \textbf{0.3795}  & \textbf{0.3209}  & \textbf{0.2899}  & \textbf{0.2710} \\\hline
    \bottomrule
    \end{tabular}
\end{adjustbox}
\caption{Effect of evidence diversification methods in performance}
\label{tab:evidence-diversification-e2e}
\end{table*}

\subsubsection{Increasing number of evidence documents}\label{sec:increase-context}

\begin{figure}[]
\centering
\includegraphics[scale=0.5]{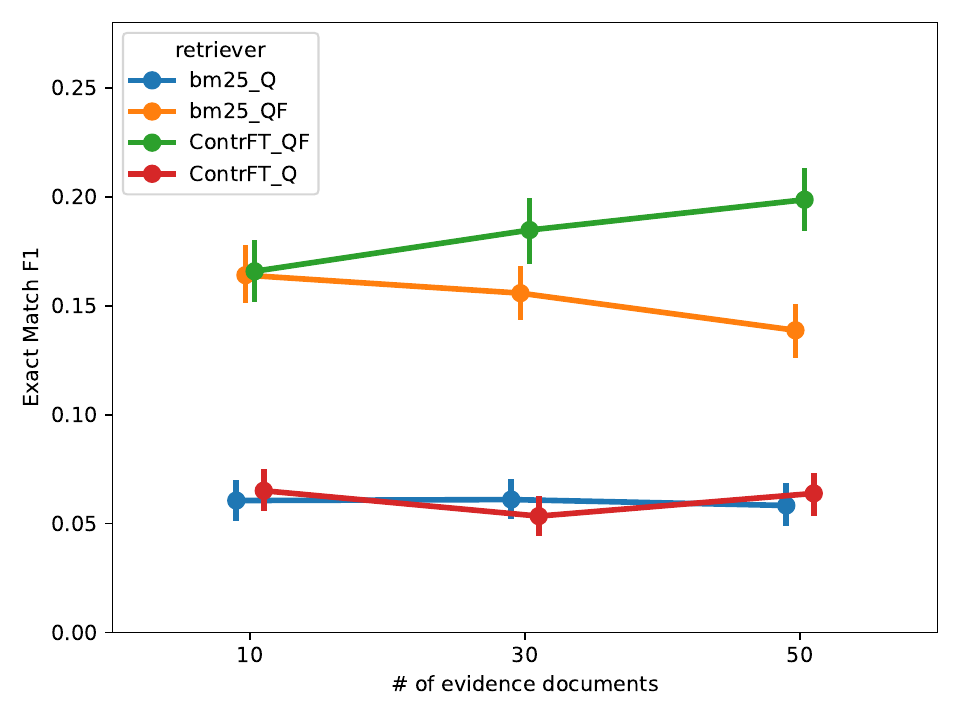}
\caption{Effect of increasing the number of evidence documents in performance}\vspace{-2em}\label{fig:n_evidence}
\end{figure} %

One of the main advantages of Fusion-in-Decoder models over other encoder-decoder models is that they can take into account lengthier context when generating a response. In practice FiD models can be viewed as compressed encoder-decoders, which can model more evidence documents with the same GPU memory footprint.
Here, we explore whether FiD can benefit from modelling more evidence documents when generating questions.
Due to computational restrictions, in this experiment we reduce the Generation Length to $32$ tokens instead of $64$ and skip the question prediction using $FiD-AspGen$. 
In Figure \ref{fig:n_evidence}, we observe that increasing the number of context documents leads to performance improvements only for $Contriever|Q,F$. 
This shows that the question generator can be improved with more documents, as long as noise is minimised in the evidence pool. This requires not only that retrieval is bounded towards the facets to be generated, but also that the retriever is of high quality (Contriever finetuned on the ranking task). For instance, performance quickly deteriorates when adding more $BM25|Q,F$ documents, probably due to adding irrelevant documents to the pool.
All in all, these results strengthen our motivation to use \textit{Fusion-in-Decoder} models for this task, since they can effectively and efficiently model larger parts of the corpus than other seq2seq counterparts. 

\textit{Section Conclusions}.
Experimental results on this Section highlight a significant gap between ground truth clarifications and the retrieval corpus in current datasets. 
When this gap appears models do not remain faithful to the facets present in the collection, ie. clarifications are not corpus-informed, 
which can severely harm search engine user experience. We show that evidence grounding can mitigate this issue and significantly boost performance.
We further verify the suitability of \textit{FiD} models for this task by showing that (i) it can effectively extract facets/topics from evidence without the need of finding them verbatim in the text, and (ii) 
its efficiency benefits can be used towards modelling larger parts of the collection and leading to generating better clarifying questions.

\subsection{Introducing novelty in the evidence pool to assist facet extraction at test time}\label{sec:results-rq3}

In previous sections, we showed that $FiD$ is effective for generating grounded Clarifying Questions end-to-end. So far, we bounded the evidence pool towards the target facets, which is not possible at inference time. 
In this Section, we explore how we can relax this condition and capture the ground truth facets in an open-domain setting.
To that end, we first examine whether grounded generators perform better than ungrounded generators in this non-bounded setting (Section \ref{sec:OOD}).
Second, we use a grounded generator that, as shown in Section \ref{sec:results-alignment} is capable to extract the facets of the evidence pool, and investigate whether diversifying the evidence pool without constraining it towards the ground-truth facets can improve results.

\subsubsection{Is evidence-facet alignment during training enough to improve facet extraction?}
\label{sec:OOD}

In Table \ref{tab:table-OOD} we use three different evidence sets at inference time and measure performance of models trained in different ways. 
We observe that differences in performance metrics are minimal within groups, while models trained and tested with evidence pools from the same evidence distribution as the test distribution is somewhat better.
This comes in contrast to the large differences observed in Table \ref{tab:evidence-effect-RAG}, where alignment brought more than double increases in Exact Match metrics, as well as Table \ref{tab:responsiveness-LOO} that showed that aligned generators are much more effective in extracting the facets present in the evidence. 
Therefore, our hypothesis remains that aligned generators do not perform better in those evidence sets because the facets do not exist there, and seek to improve this in the following Section. 
\subsubsection{Evidence diversification}\label{sec:evidence-diversification}
Our results in the previous subsection showed that grounded question generators cannot find the ground truth facets when the retrieval is not bounded towards those facets.  
In this section we explore whether introducing novelty in the evidence set can help.

To induce novelty in the inference evidence pool, we use two sets of methods. 
First, we use the Maximal Marginal Relevance (MMR) algorithm~\cite{carbonell1998use} to rerank documents from an initially retrieved evidence pool of $n=50$ documents. %
MMR reranks documents taking into account their relevance with respect to a query, but also how similar they are to the rest of the ranking list. 
In practice, this is achieved with a $\lambda$ parameter that is bounded between $0$ and $1$, where higher values of $\lambda$ give more importance to relevance than novelty.
Second, we experiment with training the dense retriever (Contriever-FT) with a diversification objective that uses knowledge distillation from the question generator~\cite{izacard2020distilling}. This method has been commonly used in Question Answering, allowing retrievers to be trained on question-answer pairs, rather than query-document relevance pairs. However, we investigate this in a different task, where we seek to optimize retrieval for novelty or diversity of retrieved passages. 
In practice, our retriever here is trained on a distillation signal from the reader, which shows which documents were attended to generate the ground truth facets. In this way, we try to bias the retriever towards retrieving the ground truth facets.

\textit{Introducing novelty using MMR.} 
The first part of Table \ref{tab:evidence-diversification-e2e} shows that inducing diversity through MMR in the evidence pool does not make the ground truth facets more likely to be retrieved and therefore extracted. Specifically, we see that increasing diversification results in lower performance.
This is consistent with the findings of \citet{samarinas2022revisiting}, that show that diversifying a list of extracted facets with MMR does not lead to improvements in MIMICS. 
Interestingly, in the related area of Multi-Answer retrieval, other works also report that MMR fails to improve results, since it increases diversity but hurts relevance significantly~\cite{min2021joint}.

\textit{Introducing novelty using Retriever-Reader Knowledge Distillation.}
In the last part of Table \ref{tab:evidence-diversification-e2e}, we experiment with finetuning the retriever jointly with the reader, using knowledge distillation from the ground truth facets. 
As usual, the retriever is initialised from the checkpoint finetuned on MSMarco. For initialising the reader, we have two different variants: $Contr-FT-Train$ is initialised directly from the Atlas checkpoint, pre-trained with the unsupervised Masked Language Modelling objective, while $Contr-FT-Train-WarmedUp$ is initialised from $[Contr-FT|Q]$.
We observe that directly finetuning a reader and retriever from the Atlas checkpoint has a detrimental effect on performance, which may be because the reader has to first learn how to adapt the generation task, from the unsupervised Masked Language Modelling objective used in pretraining to the task of Clarifying Question generation. 
However, when we initialise the model from $[Contr-FT|Q]$, training the retriever for diversification has a slight positive impact on certain performance metrics, particularly the semantic $Set-BERT$ and $Set-Blue$ metrics. In the rest of the metrics, performance remains at worst competitive with $[Contr-FT|Q]$, ie. the model we initialised from. 

We conclude that training the retriever with knowledge distillation from the reader can have a slight positive impact in performance, yet bringing up documents related to the ground truth facets remains a challenge in this dataset and task.

\section{Analysis and Discussion}\label{sec:discussion}

\subsubsection*{Implications of Taxonomy Bias}

Our preliminary qualitative analysis suggested that certain query-facet patterns are heavily repeated within the MIMICS dataset. 
This can be attributed to the dataset construction process, 
during which facets were extracted based on search engine query logs and a taxonomy defined by the authors.
To detect such patterns, we extract the top-20 frequent facet words (excluding stop-words) and inspect a couple of queries containing them. 
Based on those, we report the most frequent detected patterns in Table \ref{tab:reoccuring-patterns} and try to quantify to what extent this taxonomy bias affects the dataset. 
Our conservative estimate suggests that at least $1/5$-th of dataset queries are biased towards the predetermined taxonomy. 
We arrive to this conclusion given that roughly $20\%$ of dataset queries contain facets from this incomplete taxonomy (defined just on the top-20 facet keywords).

The implications of this for evaluation are important. 
Our experiments and related work, show that model performance has an upper-bound of $5\%-15\%$ in Exact Match metrics (depending on the type of Retrieval Augmentation). 
Yet, $~20\%$ of the dataset is (at least to some extent) described by a fixed taxonomy. 
In such a setting, it is reasonable to assume that models over-fitting to the patterns of Table \ref{tab:reoccuring-patterns} might outperform models that produce more diverse but equally reasonable query facets.
Further, models trained on this dataset are very likely to be biased towards the underlying taxonomy rather than being \textit{Corpus-informed}, raising questions regarding their ability to generalize to open-domain settings. 

This, in combination with the finding that many of the facets are irretrievable (Table \ref{tab:alignment-stats}) suggests that current test collections and evaluation frameworks are not suitable for evaluating \textit{Corpus-Informed} clarifying question generation.

\begin{table}[]
\begin{adjustbox}{width=0.95\columnwidth, center}
\begin{tabular}{l|l}
\textbf{Re-occurring patterns}& \textbf{Example}\\ 
\textit{query type}: [facets] & \\ \hline

\begin{tabular}[c]{@{}l@{}}\textit{software-related}:\, \\ {[}operating system{]}  \end{tabular}   & \begin{tabular}[c]{@{}l@{}}\textit{media creation tool}: \\ windows {[}10,7,8{]} \end{tabular}    \\ \hline

\begin{tabular}[c]{@{}l@{}}\textit{shopping query}:\, {[}review, manual, \\ specs, for sale{]}\end{tabular}         & \begin{tabular}[c]{@{}l@{}}\textit{huawei phones}:  {[}manual, specs, \\ battery, review, accessories{]}\end{tabular}  \\ \hline

\begin{tabular}[c]{@{}l@{}}\textit{location}:\, {[}hotels, \\ restaurants, population{]}\end{tabular}                    & \begin{tabular}[c]{@{}l@{}}\textit{jersey city}: \, {[}restaurants, hotels, \\ population, homes for sale{]}\end{tabular} \\ \hline

\begin{tabular}[c]{@{}l@{}}\textit{location}: \, {[}zip code, \\ weather, things to do{]}\end{tabular}             & \begin{tabular}[c]{@{}l@{}}\textit{leiden}: \, {[}things to do, weather,\\  zip code, what time is it{]}\end{tabular}     \\ \hline

\begin{tabular}[c]{@{}l@{}}\textit{movie} or \textit{series}: \, {[}cast, trailer, \\ quotes, review{]}\end{tabular}       & \begin{tabular}[c]{@{}l@{}}\textit{her alibi movie}:\, {[}cast, trailer,\\ review, quotes{]}\end{tabular}         \\ \hline

\begin{tabular}[c]{@{}l@{}}\textit{medical condition}: \, {[}symptom,\\ treatment, causes, diagnosis{]}\end{tabular} & \begin{tabular}[c]{@{}l@{}}\textit{adhd}: \, {[}symptom, treatment, \\ causes, diagnosis, diet{]}\end{tabular}            \\ \hline

\begin{tabular}[c]{@{}l@{}}\textit{act}: \,  {[}checklist, tips, hacks{]}\end{tabular}                      & \begin{tabular}[c]{@{}l@{}}\textit{getting a kitten}: \, {[}tips, checklist, \\ life hacks, pros{]}\end{tabular} \\\hline

\textit{various}:\,   {[}men, women, kids{]}& \begin{tabular}[c]{@{}l@{}}\textit{short layered hairstyles}: \\ {[}women, men, teens, kids{]}\end{tabular}           \\\hline

\multicolumn{2}{c}{Dataset percentage: $20.74\%$} \\\hline

\end{tabular}

\end{adjustbox}
\caption{Reoccurring facet patterns in the MIMICS dataset}\label{tab:reoccuring-patterns}
\end{table}

\subsubsection*{Qualitative analysis}\vspace{-0.5em}
In Table \ref{tab:query-analysis} we analyse a handful of random queries, presenting their ground-truth and generated facets. 
We show facets generated by our aligned model (train. evidence: \textit{Contr-FT|Q,F}), 
and tested in an open-domain setting without diversification (test evidence: \textit{Contr-FT|Q})(second-to-last row of Table \ref{tab:table-OOD}). 
Firstly, we observe that even the model trained with alignment learns to reproduce common re-occurring facet patterns from the training set (eg. ``windows 7,8,10''). 
In fact, this still happens when evidence documents are unrelated to those facet patterns, as only 2/10 evidence documents mention operating systems versions, while the topic of the documents is not specifically related to those. 
Another observation that stands out is that the ground-truth facets are not exhaustive and other equally good or better facets can be generated. Such examples include \textbf{episodes} of penny appearing in the big bang theory, or referring to the leiden \textbf{thrombophilia} disorder rather than aspects related to the city.

Given those static ground truths, evaluation metrics would be very low, despite the overall high quality. 
For the queries ``network drive'', ``leiden'', ``sickle cell anemia'' and ``consumer price index'', both Precision and Recall would be zero, despite the quality and greater diversity of the produced facets. 
This highlights the the presence of \textbf{multiple ground truths}, which are in fact ignored in the current evaluation framework.

This analysis highlights an important gap in current datasets and evaluation protocols, that prohibits the generation of \textit{Corpus-informed} clarifying questions:
Ground truth facets are constructed from query reformulations and static taxonomies, but do not reflect the facets that are often present in search collections.

\begin{table}[]
\begin{adjustbox}{width=0.95\columnwidth, center}
\begin{tabular}{l|ll}
\textbf{query}                                                            & \begin{tabular}[c]{@{}l@{}}\textbf{facets}\\ \textbf{(ground truth)}\end{tabular}                                                       & \begin{tabular}[c]{@{}l@{}}\textbf{facets}\\ \textbf{(generated)}\end{tabular}               \\\hline\hline

network drive & set up, remove& \begin{tabular}[c]{@{}l@{}}windows 7, windows 8,\\ windows 10\end{tabular} \\\hline

leiden& \begin{tabular}[c]{@{}l@{}}what time is it, zip code,\\ things to do, weather\end{tabular}
& \textbf{leiden thrombophilia} \\\hline

\begin{tabular}[c]{@{}l@{}}penny big \\ bang theory\end{tabular} & cast, quotes& \underline{dvd}, cast, \textbf{episodes}\\\hline

pemetrexed & \begin{tabular}[c]{@{}l@{}}side effects, chemotherapy\\ injection, mechanism of \\ action, package insert\end{tabular} & side effects, \textbf{cost}\\\hline

\begin{tabular}[c]{@{}l@{}}sickle cell \\ anemia\end{tabular}    & \begin{tabular}[c]{@{}l@{}}diagnosis, causes,\\ diet, symptom, treatment\end{tabular}                                 & \begin{tabular}[c]{@{}l@{}}\textbf{in children},\\ \textbf{in adults}\end{tabular}\\\hline

\begin{tabular}[c]{@{}l@{}}consumer price \\ index\end{tabular}    & U.S., india, japan & \textbf{chart, calculator}

\end{tabular}
\end{adjustbox}
\caption{Qualitative analysis. Generated facets with an aligned question generator (train. evidence: \textit{Contr-FT|Q,F}) in an open domain setting (test evidence: \textit{Contr-FT|Q}).
\underline{Underlined} indicate irrelevant facets, and 
\textbf{boldfaced} indicate high-quality facets outside of the ground truth.}\label{tab:query-analysis}
\end{table}

\section{Conclusions and future directions}

In this paper, we investigated the task of generating \emph{Corpus-Informed} Clarifying Questions end-to-end, based on Retrieval Augmented Generation models. 
We showed that \textit{Fusion-in-Decoder} models are able to effectively and efficiently model queries and evidence documents when generating clarifying questions. %
This efficiency advantage allows them to model larger parts of the corpus when asking clarification questions, potentially improving question quality.

Further, we investigated the role of retrieval in this task, showing that retrieval quality is more important for the generator than finding exact lexical matches of the facets to be generated.
We showed that current datasets suffer from a lack of grounding between ground truth facets and evidence documents, which has serious implications.%
Specifically, it hinders their ability of models to ask \textit{Corpus-informed} questions, making them prone to "hallucinations" (ie. ignoring evidence documents while generating facets), which can severely harm user experience. 

Lastly, we attempt to construct an evidence pool that contains the ground truth facets by inducing novelty in the retriever.
We experiment with two novelty methods, (i) $MMR$ and (ii) training the retriever with knowledge distillation, but we conclude that doing so remains a challenge. 
Our analysis sheds light in those challenges and highlights the gap between ground truth facets and those present in existing ranking corpora, calling for the need of datasets that can support generating Corpus-Informed clarifications.
Future work should focus in developing appropriate train and test collections that better reflect the objective of creating \textit{Corpus-Informed} Clarifying Questions.  %
Novelty and diversity of the evidence pool remains important for generating good questions. To this end, exploring probabilistic retrieval methods~\cite{warburg2023bayesian,zamani2023multivariate}, 
and combining them with retriever training techniques, such as reader distillation remains a promising direction.

\clearpage
\newpage
\bibliographystyle{ACM-Reference-Format}
\balance
\bibliography{main}

\end{document}